\documentclass[12pt]{article}
\usepackage{pic03}
\usepackage{hyperref}
\usepackage{url}
\usepackage{graphicx}
\def\Missing#1#2{{\mbox{$#1\kern-0.57em\raise0.19ex\hbox{/}_{#2}$}}}
\newcommand{\gevcc}{\mbox{GeV/$c^2$}}                   %GeV/c^2
\newcommand{\jjmass}{\mbox{$M_{JJ}$}}
\newcommand{\ipb}{\mbox{pb$^{-1}$}}
\newcommand{\pbarp}{\mbox{$p\overline{p}$}}
\newcommand{\pt}{\mbox{$p_{T}$}}
\newcommand{\Et}{\mbox{$E_{T}$}}
\newcommand{\mody}{\mbox{$\mid \! y  \! \mid$}}
\newcommand{\modeta}{\mbox{$\mid \! \eta  \! \mid$}}
\newcommand{\modz}{\mbox{$\mid \! z  \! \mid$}}
\newcommand{\met}{\mbox{$\Missing{E}{T}$}}

\newcommand{\ptu}[1]{\mbox{$p^{\rm jet#1}_{T}$}}
\newcommand\prl[3]{{Phys.\ Rev.\ Lett.\ }{\bf #1} (#2) #3}
\newcommand\npb[3]{{Nucl.\ Phys.\ }{\bf B #1} (#2) #3}
\newcommand\prd[3]{{Phys.\ Rev.\ }{\bf D #1} (#2) #3}

\newcommand{\cteqfourhj}{\mbox{CTEQ4HJ}}

\begin{document}

\title{\bf Jet Physics at the Tevatron}
\author{
Iain A. Bertram        \\
{\em Lancaster University}}
\maketitle

%
% photograph of author
%  This is where we will insert a photograph. To see what it would look like,
%  uncomment the following lines.
%
%\begin{figure}[h]
%\begin{center}
%
% include photograph for proceeding version
%
%\includegraphics
%[height=4.5cm]{einstein.eps}
%
% insert a fixed vertical spacing instead for the ArXiv preprint
%
\vspace{4.5cm}
%
%\end{center}
%\end{figure}

\baselineskip=14.5pt
\begin{abstract}
 Results are presented from analyses of jet data produced in \pbarp\
 collisions at $\sqrt{s}$ = 1960 GeV collected with the D\O\ and CDF
 detectors during 2002--03 at the Fermilab Tevatron
 Collider. Preliminary measurements of the inclusive jet cross
 section, the dijet mass spectrum, and jet structure are presented.
\end{abstract}
\newpage

\baselineskip=17pt

\section{Introduction}

 Perturbative QCD (pQCD) predicts the production cross sections at
 large transverse momentum ($p_{T}$) for parton-parton scattering in
 proton--antiproton (\pbarp ) collisions. The outgoing partons from
 the parton-parton scattering hadronize to form jets of particles.
 Calculations of high-\pt\ jet production involve the folding of
 parton scattering cross sections with experimentally determined
 parton distribution functions (PDFs).  These predictions are
 calculated to next-to-leading-order (NLO) QCD
 calculations~\cite{aversa,eks,jetrad}.  In this paper I present
 several measurements of jet cross sections at $\sqrt{s}$ = 1960 GeV
 collected with the D\O\ and CDF detectors. 

 In the analyses presented in this paper Jets are reconstructed using
 an iterative cone algorithm, known as the improved legacy cone
 algorithm~\cite{ilca}, with a fixed cone radius of $\mathcal{R}=0.7$
 in $\eta$--$\phi$ space, where $\varphi$ is the azimuth. This cone
 algorithm clusters jets about seeds. If two seeds are within
 $2\mathcal{R}$ of each other, a third seed is created at the midpoint
 between them. Jets with overlapping cones are merged or split
 according to the following criteria: two jets are merged into one jet
 if more than 50$\%$ of the $\Et$ of the jet with the smaller $\Et$ is
 contained in the overlap region.  If less than 50$\%$ of the $\Et$ is
 contained in the overlap region, the jets are split into two distinct
 jets and the energy of each calorimeter cell in the overlap region is
 assigned to the nearest jet.  The jet directions are then
 recalculated.

 The analyses presented in this paper were carried out using data
 collected during 2002-2003. The D\O\ data sample corresponds to an
 integrated luminosity of 34~\ipb\ and the CDF data sample 85\ipb .

\section{Run 2 at the Tevatron}

 The Tevatron has recently been upgraded to improve both the
 luminosity and the center-of-mass (CM) energy. During the shutdown
 between Run 1 (1992--1995) and the current Run 2 (2002--), the
 accelerator's Main Ring was replaced with the Main Injector which
 should provide higher beam currents and cycling rates. .  The CM
 energy has been increased from $\sqrt{s} = 1800$~GeV in Run~1 to
 $\sqrt{s} = 1960$~GeV.

 To cope with the increased luminosity the D\O\ and CDF experiments
 were both upgraded during the shutdown period. The D\O\ detector has
 upgraded its tracking system, with the addition of a 2T Solenoid
 magnet surrounding a silicon microstrip tracker, and a scintillating
 fiber tracker. The calorimeter's readout electronics have been
 replaced, and central and forward preshower detectors have been added
 between the solenoid and the calorimeter. The muon detector system
 has been extended by adding scintillating layers for triggering,
 extending the drift chamber coverage, and improving the beamline
 shielding. The entire trigger and DAQ system was also replaced to
 cope with the increased beam crossing rate.

 The CDF detector underwent a similar upgrade. The tracking systems
 were replaces with a new silicon microstrip tracker, Time-of-flight
 detector, and drift chamber. The calorimeter coverage was improved by
 adding new plug and mini-plug calorimeters. The muon coverage was
 also improved. Finally, CDF also upgraded their entire trigger and
 DAQ system as well.

\section{Inclusive Jet Cross Section}

\begin{figure}[htb]
  \centerline{
    \parbox{.475\textwidth}{\centerline{\includegraphics[width=.475\textwidth ]{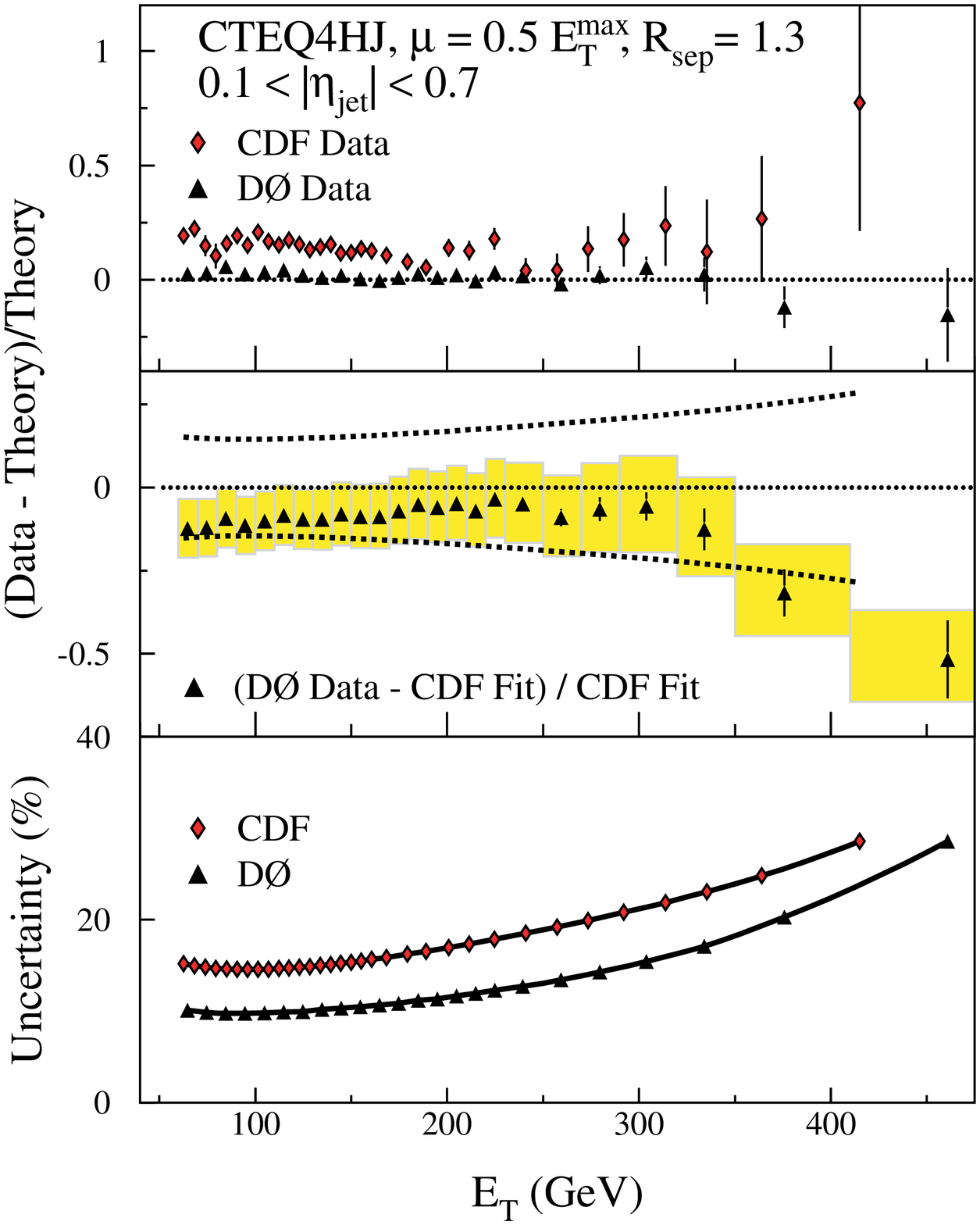}
    }}~~~~
    \parbox{.475\textwidth}{\centerline{\includegraphics[width=.475\textwidth ]{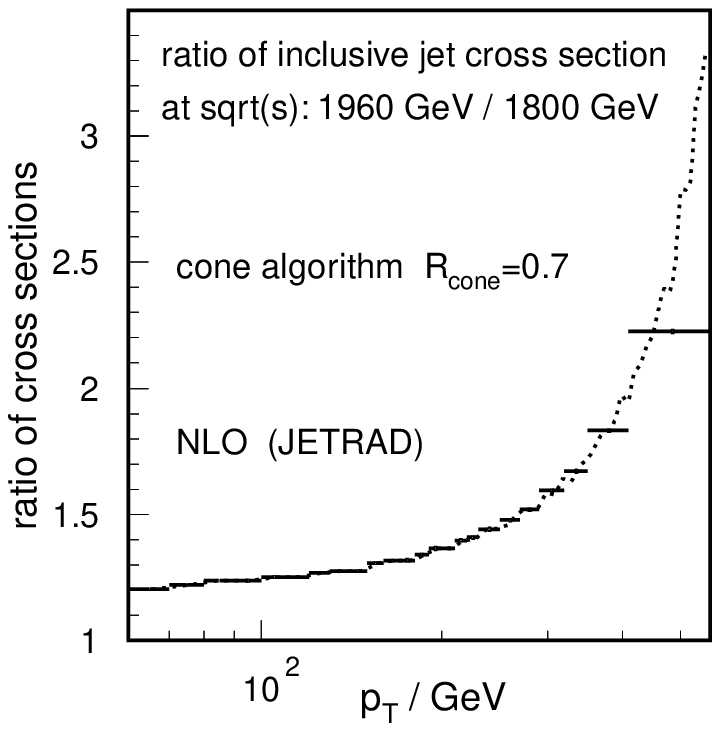}
    }}
  }
  \centerline{
    \parbox[t]{.475\textwidth}{\caption{Inclusive Jet Cross Sections
	for $0.1 < \modeta < 0.7$ from D\O\ and CDF compared to the
	theory prediction {\sc Jetrad} with the CTEQ4HJ
	PDF. \label{inc_comp}} }~~~~
    \parbox[t]{.475\textwidth}{\caption{The ratio of predicted
      inclusive jet cross sections at $\sqrt{s}=1960$~GeV and
      $\sqrt{s}=1800$~GeV for $\modeta < 0.5$ using {\sc Jetrad} with
      the CTEQ4HJ PDF.
      \label{run2_pred}} } }
\end{figure}

 The high \pt\ behaviour of the inclusive jet cross section has been
 the subject of much discussion. The previous measurements of the
 cross section by D\O ~\cite{d0prd} and CDF~\cite{cdfinc} are compared
 with the {\sc Jetrad}~\cite{jetrad} NLO Monte Carlo prediction using
 the \cteqfourhj\ PDF in Fig.~\ref{inc_comp}. The CDF measurement
 showed a possible excess of high-\pt\ jet production, while the D\O\
 measurement is in good agreement with the predictions. The two
 measurements are statistically consistent with each
 other~\cite{d0prd}. The slight excess can be explained using PDFs
 that include the Run 1 high-\pt\ jet data (CTEQ6~\cite{cteq6} and
 MRST2001~\cite{mrst2001}) or by the introduction of new phyiscs
 beyond the standard model. 

 The increase in CM energy from $\sqrt{s}=1800$ to $\sqrt{s}=1960$~GeV
 means that the cross section will increase by $40\%$ (200\%) for jets
 of 300 (400) GeV (Fig.~\ref{run2_pred}). The Tevatron is expected to
 deliver 2~\ipb\ during the first phase of Run~2 compared with a total
 data sample of approximately 110~\ipb\ in Run 1. This will result in
 a dramatic increase of statistics that will result in measurements
 that are dominated by systematic uncertainties for almost the entire
 \pt\ range measured.

\begin{figure}[htb]
  \centerline{
    \parbox{.475\textwidth}{\centerline{\includegraphics[width=.475\textwidth ]{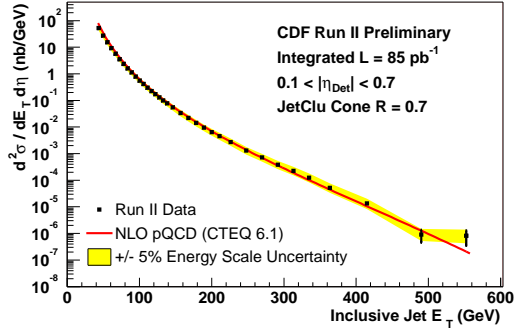}
    }}~~~~
    \parbox{.475\textwidth}{\centerline{\includegraphics[width=.475\textwidth ]{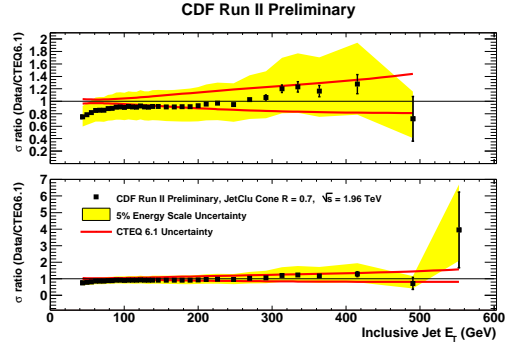}
    }}
  }
  \centerline{
    \parbox[t]{\textwidth}{\caption{The inclusive jet cross section as
    measured by CDF for $0.1 < \mody < 0.7$ compared with the {\sc
    Jetrad} prediction with the CTEQ61 PDF.  The left hand plot is on
    a logarithmic scale and the right hand plot shows the data divided
    by the theoretical prediction.
    \label{cdf_inc1}} }
    }
\end{figure}

 CDF has measured the inclusive jet cross section using a data sample
 of 85~\ipb\ in the rapidity range $0.1 < \mody < 0.7$. The data was
 collected using four triggers with uncorrected jet transverse energy
 (\Et ) thresholds of 20, 50, 70, and 100~GeV. The $z$-position of the
 event vertex was required to satisfy $\mid \! z \! \mid < 60$cm, and
 the effects of noise and cosmic rays were removed using cuts on the
 missing-\Et\ and by event scanning. The resulting cross section was
 corrected for energy scale effects and unsmeared. The uncertainty on
 the jet \Et\ due to the jet energy scale correction is currently
 $5\%$. The resulting cross section is then compared with a {\sc
 Jetrad} prediction with the CTEQ61 PDF~\cite{cteq6}
 (Fig.~\ref{cdf_inc1}). The data show good agreement with the
 theoretical predictions given the size of the experimental
 uncertainties.

\begin{figure}[htb]
  \centerline{
    \parbox{.475\textwidth}{\centerline{\includegraphics[width=.475\textwidth ]{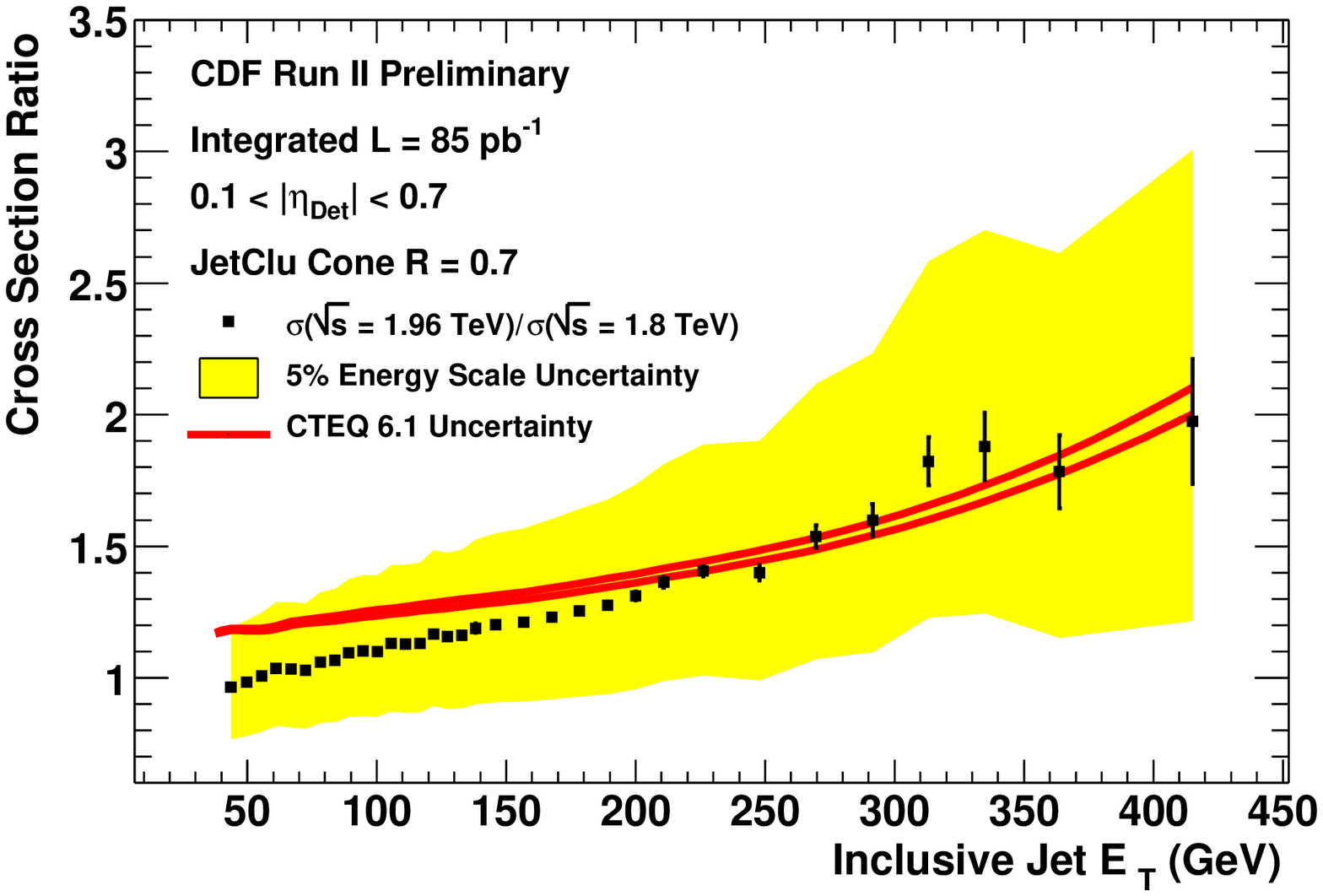}
    }}~~~~
    \parbox{.475\textwidth}{\centerline{\includegraphics[width=.475\textwidth ]{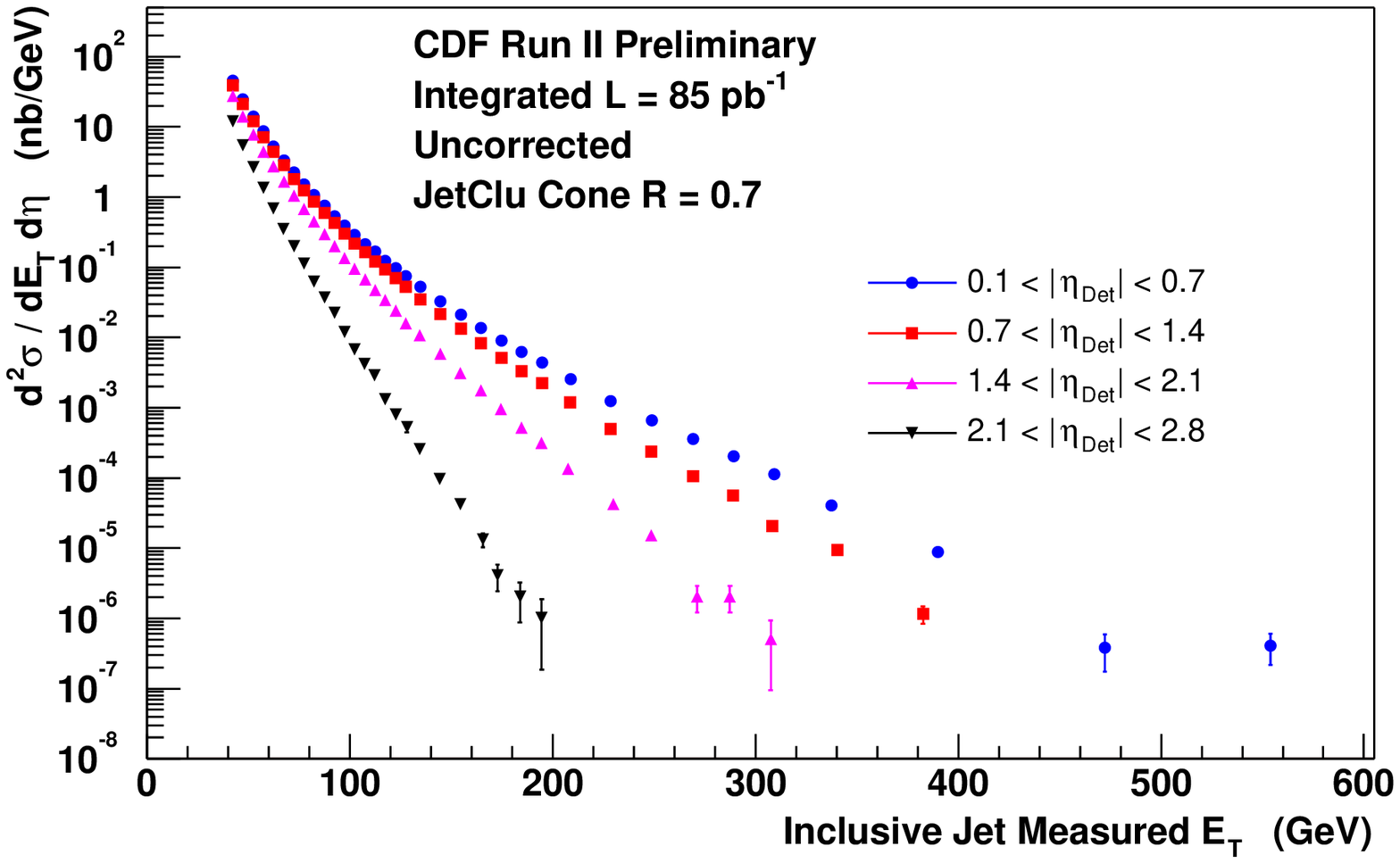}
    }}
  }
  \centerline{
    \parbox[t]{.475\textwidth}{\caption{The ratio of inclusive jet
    cross sections at $\sqrt{s}=1960$ divided by $\sqrt{s}=1800$~GeV
    by CDF for $0.1 < \mody < 0.7$ compared with the {\sc Jetrad}
    prediction. \label{cdf_inc3}} }~~~~
    \parbox[t]{.475\textwidth}{\caption{The CDF inclusive jet cross
    section for $0.1 < \mody  0.7$, $0.7 < \mody < 1.4$, $1.4 <
    \mody < 2.4$, and $2.1 < \mody < 2.8$.
      \label{cdf_inc4}} } }
\end{figure}

 CDF has carried out a comparison between the inclusive jet cross
 section measured at $\sqrt{s}=1800$ and $\sqrt{s}=1960$~GeV
 (Fig.~\ref{cdf_inc3}). The data are in good agreement with the
 theoretical predictions for jet-\Et\ greater than 250~GeV. At lower
 values of \Et\ the Run~2 data show an excess compared with the
 expected cross section. Increased statistics and improved
 understanding of systematic uncertainties over the next year should
 shed some light on this.

 CDF has also measured the jet cross section out to an $\eta$ of 2.8
 for the first time (Fig.~\ref{cdf_inc4}). These measurements will add
 to the current D\O\ Run~1 measurement~\cite{d0_forward} of the jet
 cross section at forward rapidities as an important component of
 future PDF fits.

\begin{figure}[htb]
  \centerline{
    \parbox{.475\textwidth}{\centerline{\includegraphics[width=.475\textwidth ]{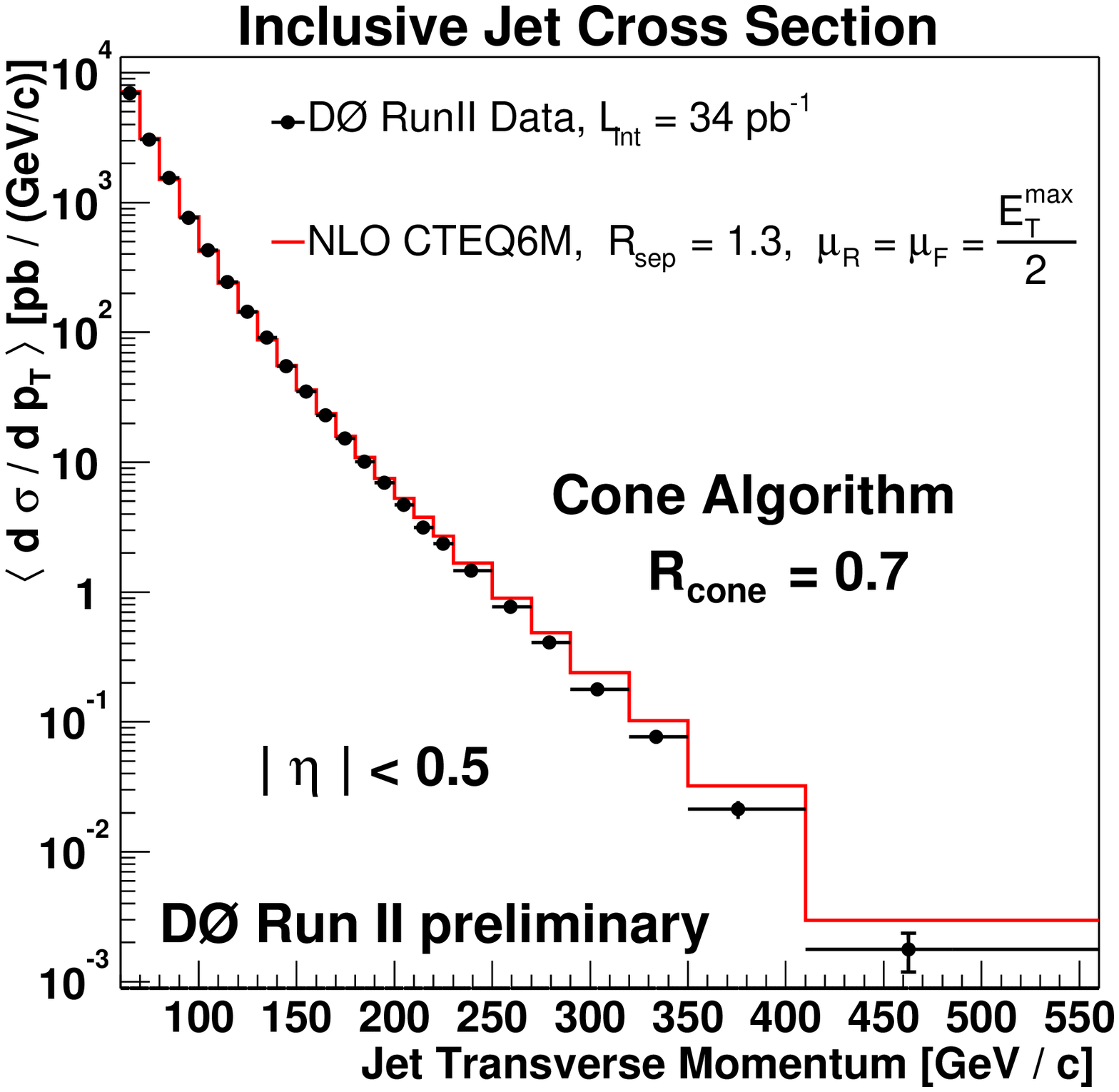}
    }}~~~~
    \parbox{.475\textwidth}{\centerline{\includegraphics[width=.475\textwidth ]{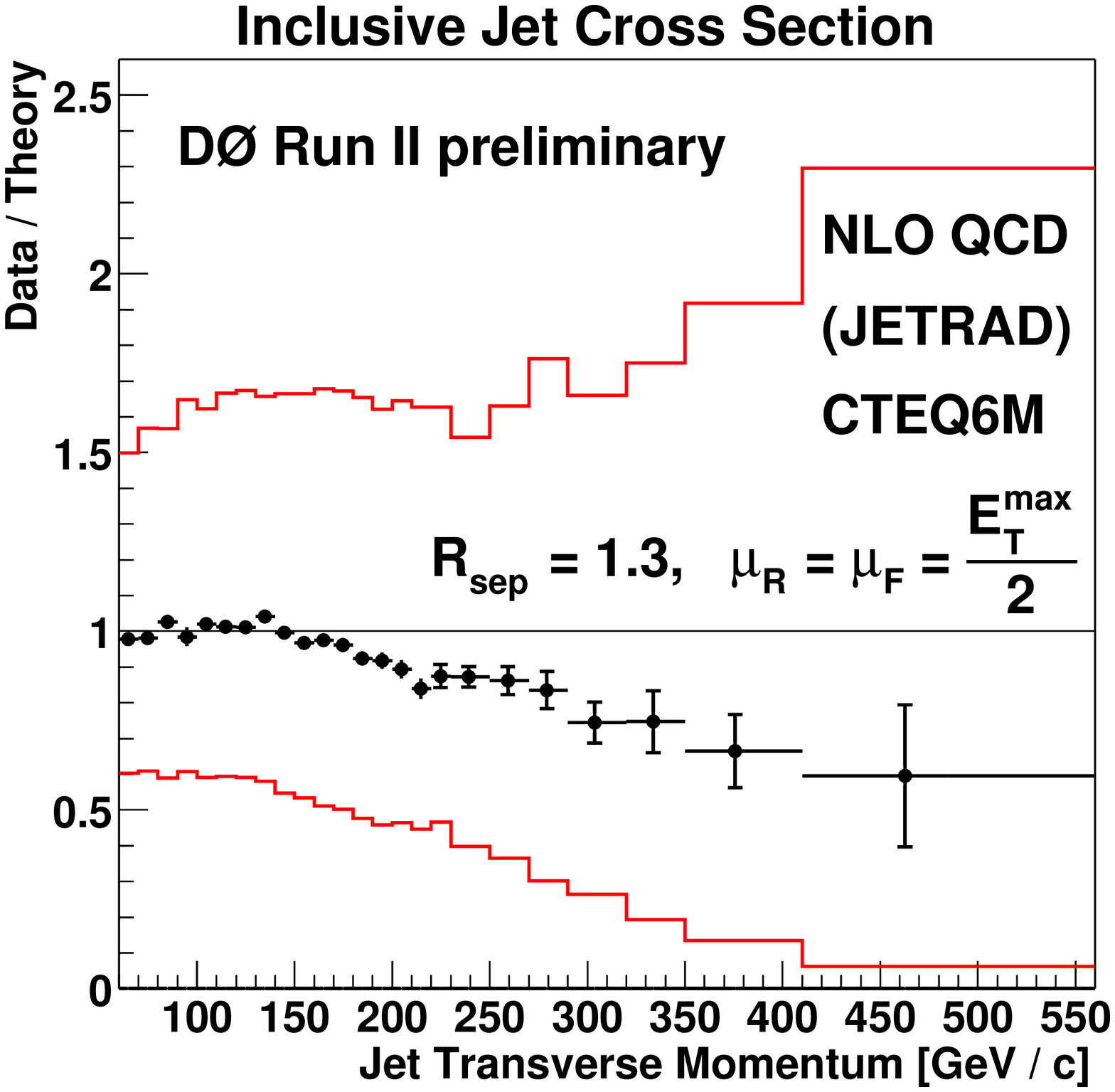}
    }}
  }
  \centerline{
    \parbox[t]{\textwidth}{\caption{The inclusive jet cross section as
    measured by D\O\ for $\mody < 0.5$ compared with the {\sc Jetrad}
    prediction with the CTEQ6M PDF.  The left hand plot is on a
    logarithmic scale and the right hand plot shows The difference
    between the data and the prediction, divided by the prediction.
    \label{d0_inc1}} } }
\end{figure}

 D\O\ measured the inclusive jet cross section using a data sample of
 34~\ipb\ in the rapidity range $\mody < 0.5$. The data were collected
 with five triggers with uncorrected-\Et\ threshold of 5, 25, 45, 65,
 and 95~GeV. The $z$-vertex was required to satisfy the cut $\modz <
 50$~cm. Jets caused by noise and cosmic rays were rejected using the
 cut: ${\met}/{\ptu{1}} < 0.7$, where \ptu{1}\ is the transverse
 momentum of the highest \pt\ jet in the event, and \met\ is the
 missing transverse energy in the event. Additional quality cuts are
 made on the jet shower shapes to eliminate any remaining noisy
 jets. The resulting cross section is then compared with a {\sc
 Jetrad} prediction with the CTEQ6M PDF~\cite{cteq6}
 (Fig.~\ref{d0_inc1}). The prediction and the measured cross section
 are in good agreement given the large size of the experimental
 uncertainties which are dominated by the uncertainty on the jet
 energy scale. The energy scale uncertainty is expected to halve by
 the end of summer 2003. At this time no comparison has been made
 between D\O\ and CDF results, and both experiments are in good
 agreement with {\sc Jetrad} predictions that use the MRST2001
 PDFs~\cite{mrst2001}.

\section{Dijet Measurements}

\begin{figure}[hbt]
  \centerline{
    \parbox{.475\textwidth}{\centerline{\includegraphics[width=.475\textwidth ]{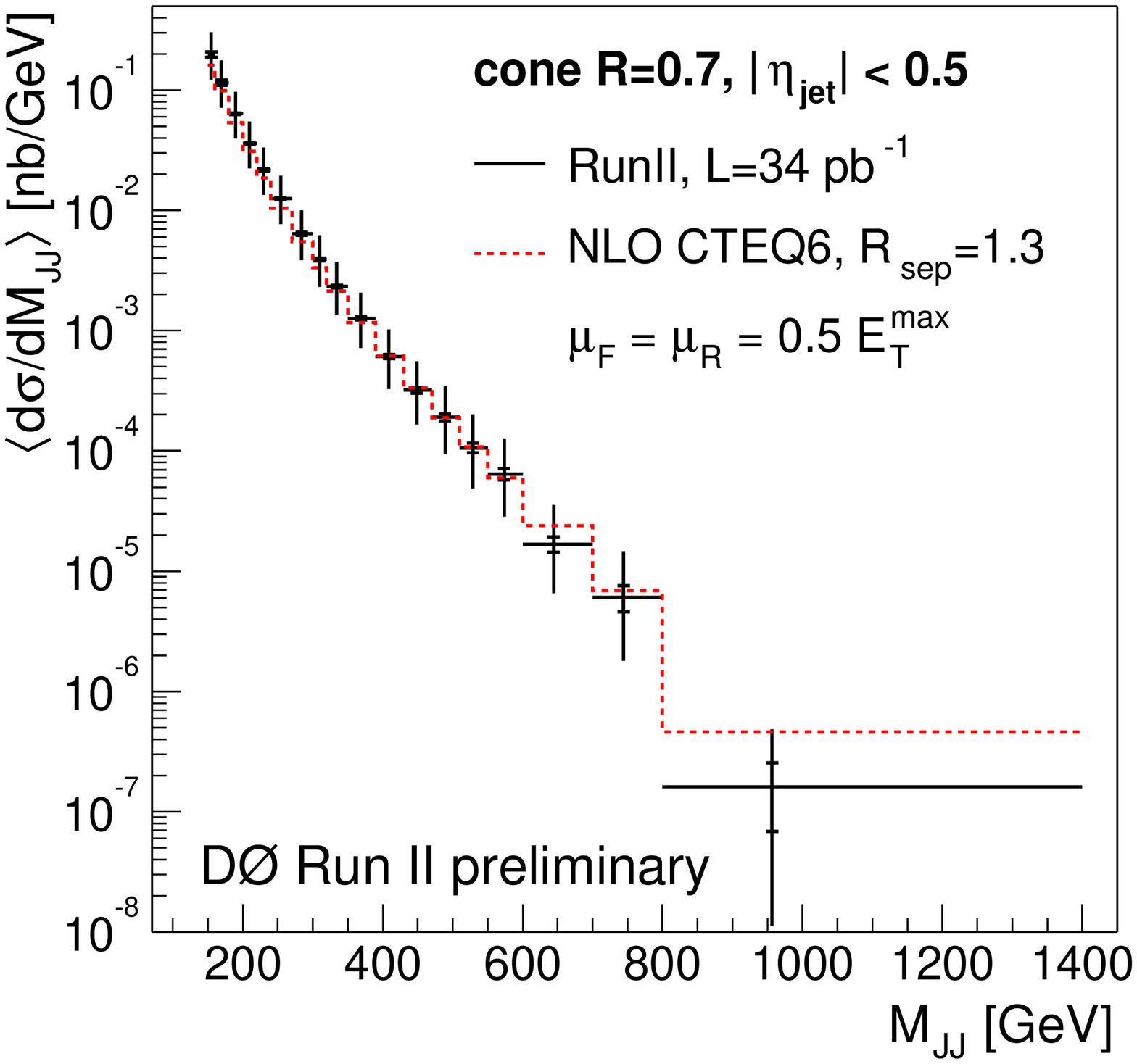}
    }}~~~~
    \parbox{.475\textwidth}{\centerline{\includegraphics[width=.475\textwidth ]{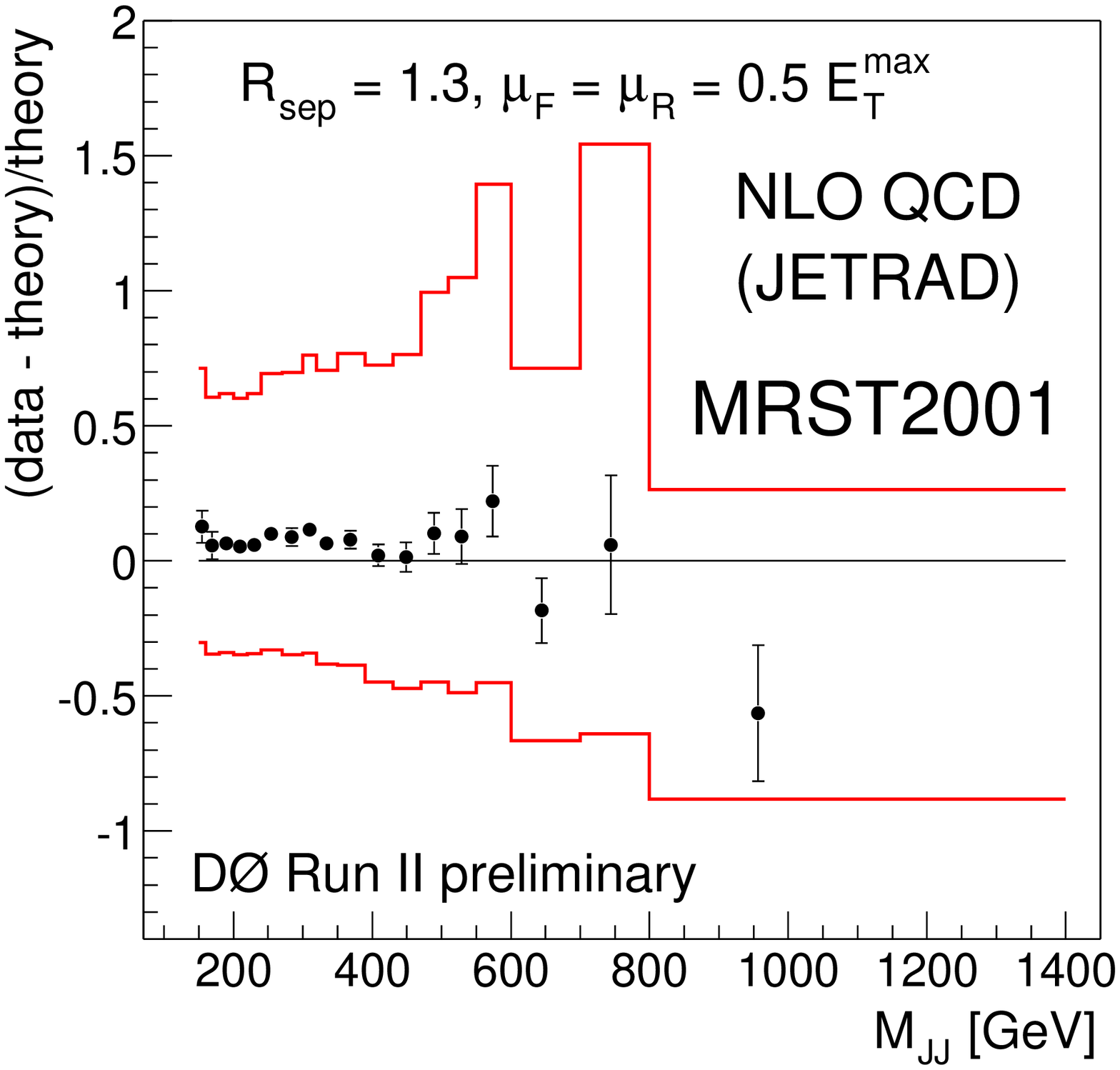}
    }}
  }
  \centerline{
    \parbox[t]{\textwidth}{\caption{The dijet mass cross section as
    measured by D\O\ for $\mody < 0.5$ compared with {\sc Jetrad}
    predictions.  The left hand plot is on a logarithmic scale and
    uses the CTEQ6M PDF and the right hand plot shows The difference
    between the data and the prediction, divided by the prediction
    using the MRST2001 PDF.
    \label{d0_mass1}} } }
\end{figure}

 D\O\ has made a measurement of the dijet mass cross section using the
 same data sample as used in the measurement of the inclusive jet
 cross section. The dijet mass is calculated using the two highest
 \pt\ jets in the event. The mass is defined as $\jjmass^2 =
 {(E_1+E_2)^2 - ({\vec{p_1}} + {\vec{p_2}})^2}$. Both jets in the
 event are required to have $\mody < 0.5$. The data and theoretical
 predictions are in good agreement (Fig.~\ref{d0_mass1}).

\begin{figure}[htb]
  \centerline{
    \parbox{.475\textwidth}{\centerline{\includegraphics[width=.475\textwidth ]{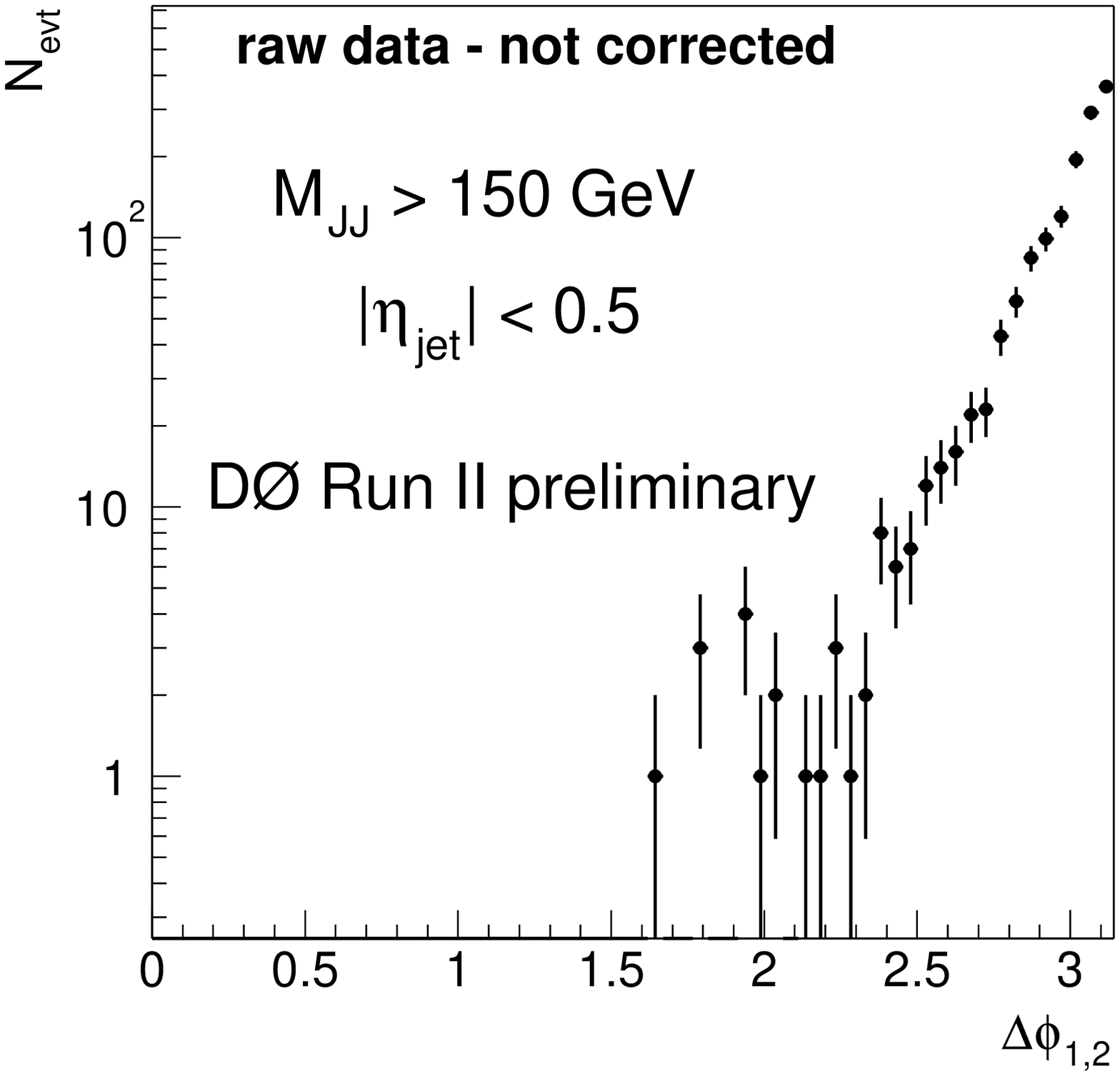}
    }}~~~~
    \parbox{.475\textwidth}{\centerline{\includegraphics[width=.475\textwidth ]{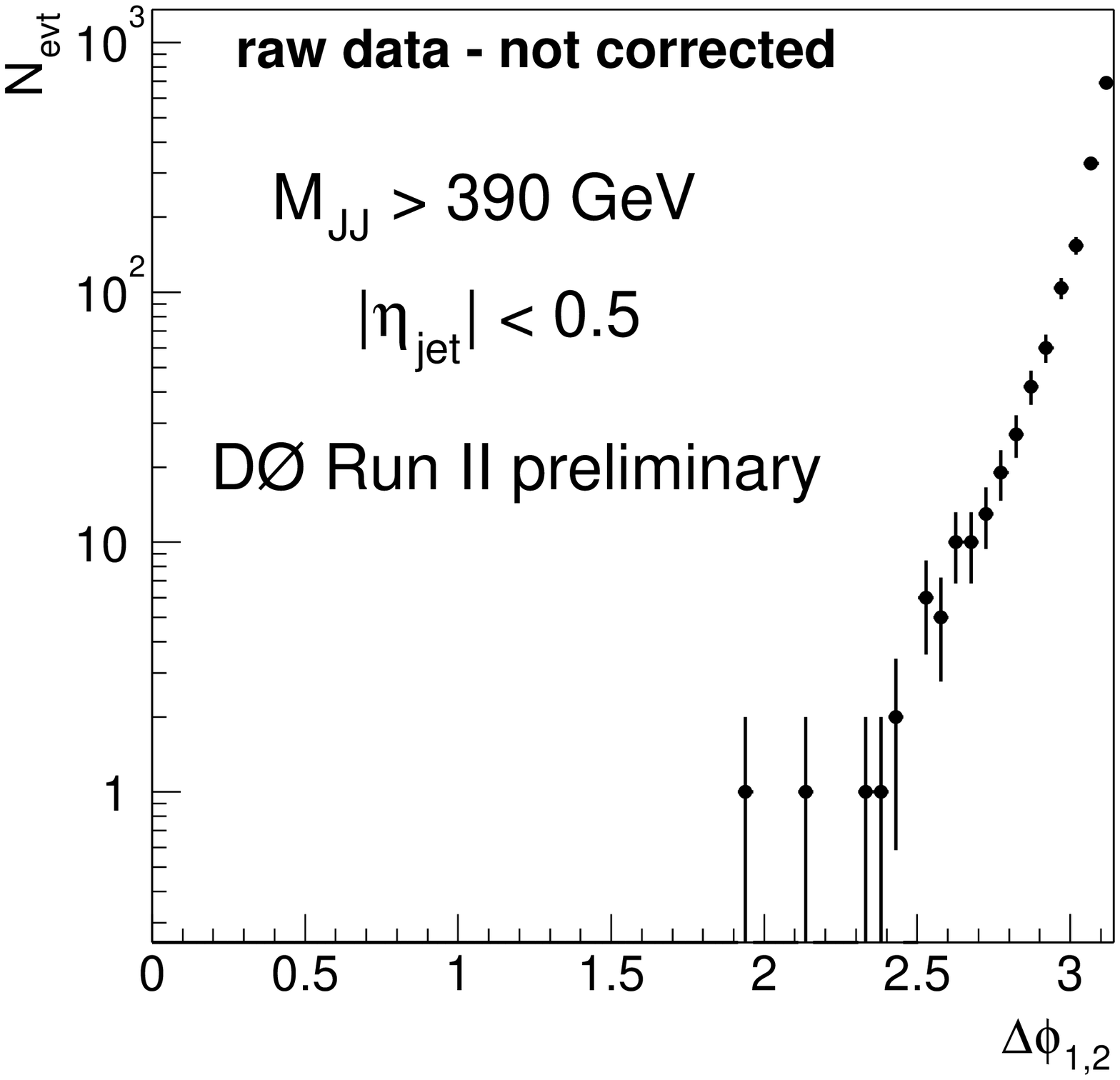}
    }}
  }
  \centerline{
    \parbox[t]{\textwidth}{\caption{The separation in $\phi$ between
    the two highest-\pt\ jets in the event above a given mass
    threshold. The left hand plot is for $\jjmass > 150$~\gevcc\ and
    the right hand plot is for $\jjmass > 390$~\gevcc .
    \label{d0_phi}} } }
\end{figure}

 In addition to measuring the mass in these events D\O\ also measured
 the separation in the azimuthal angle, $\phi$, between the two
 highest \pt\ jets in the events that satisfy $\mody < 0.5$. Four
 measurements are made, using four triggers with thresholds 25, 45,
 65, and 95~GeV corresponding to mass bins of 150, 180, 300, and
 390~\gevcc\ respectively. The resulting distributions
 (Fig.~\ref{d0_phi}) show that jets become more back-to-back as the
 mass of the events increases.

\begin{figure}[htb]
  \centerline{
    \parbox{.475\textwidth}{\centerline{\includegraphics[width=.475\textwidth ]{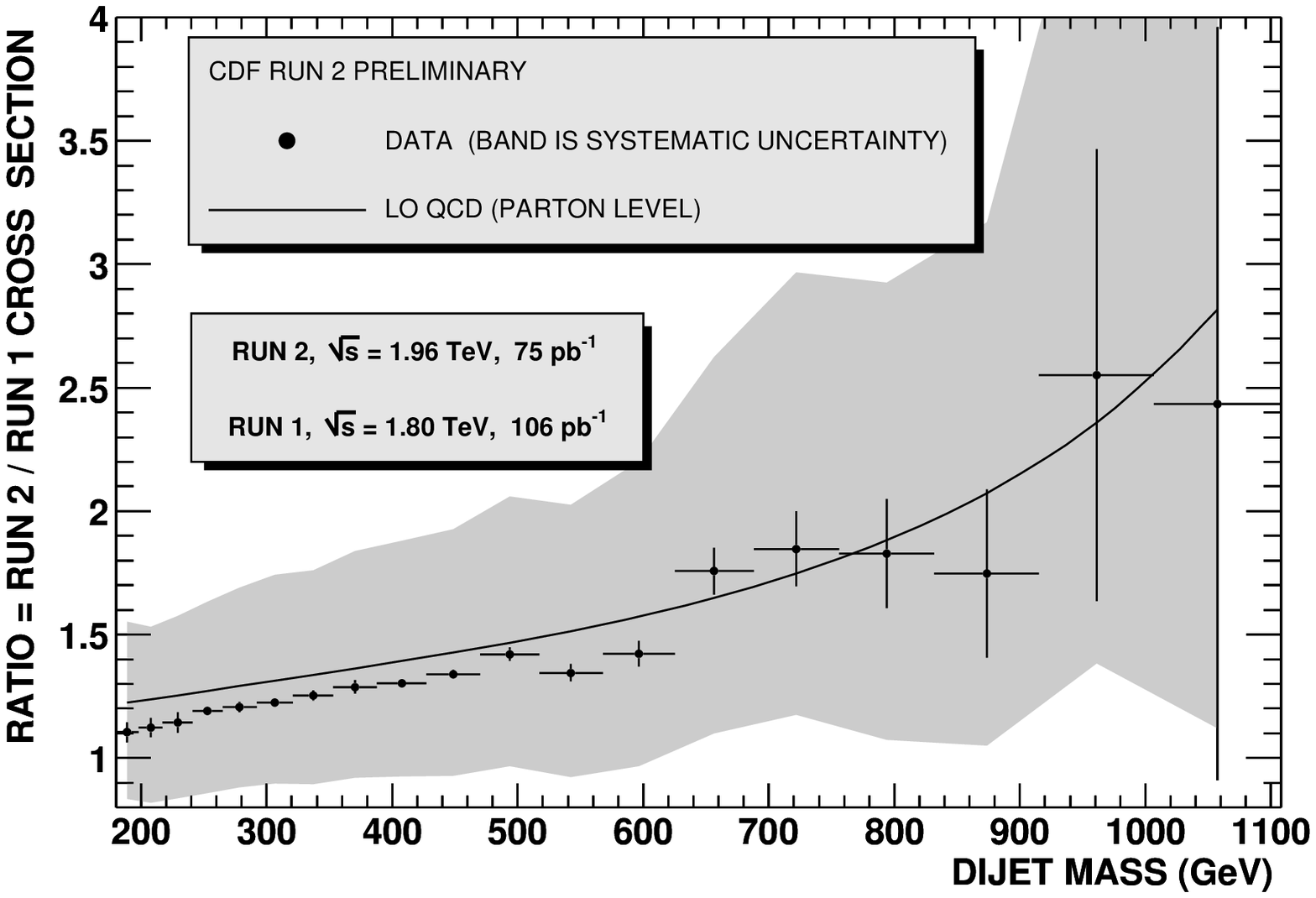}
    }}~~~~
    \parbox{.475\textwidth}{\centerline{\includegraphics[width=.475\textwidth ]{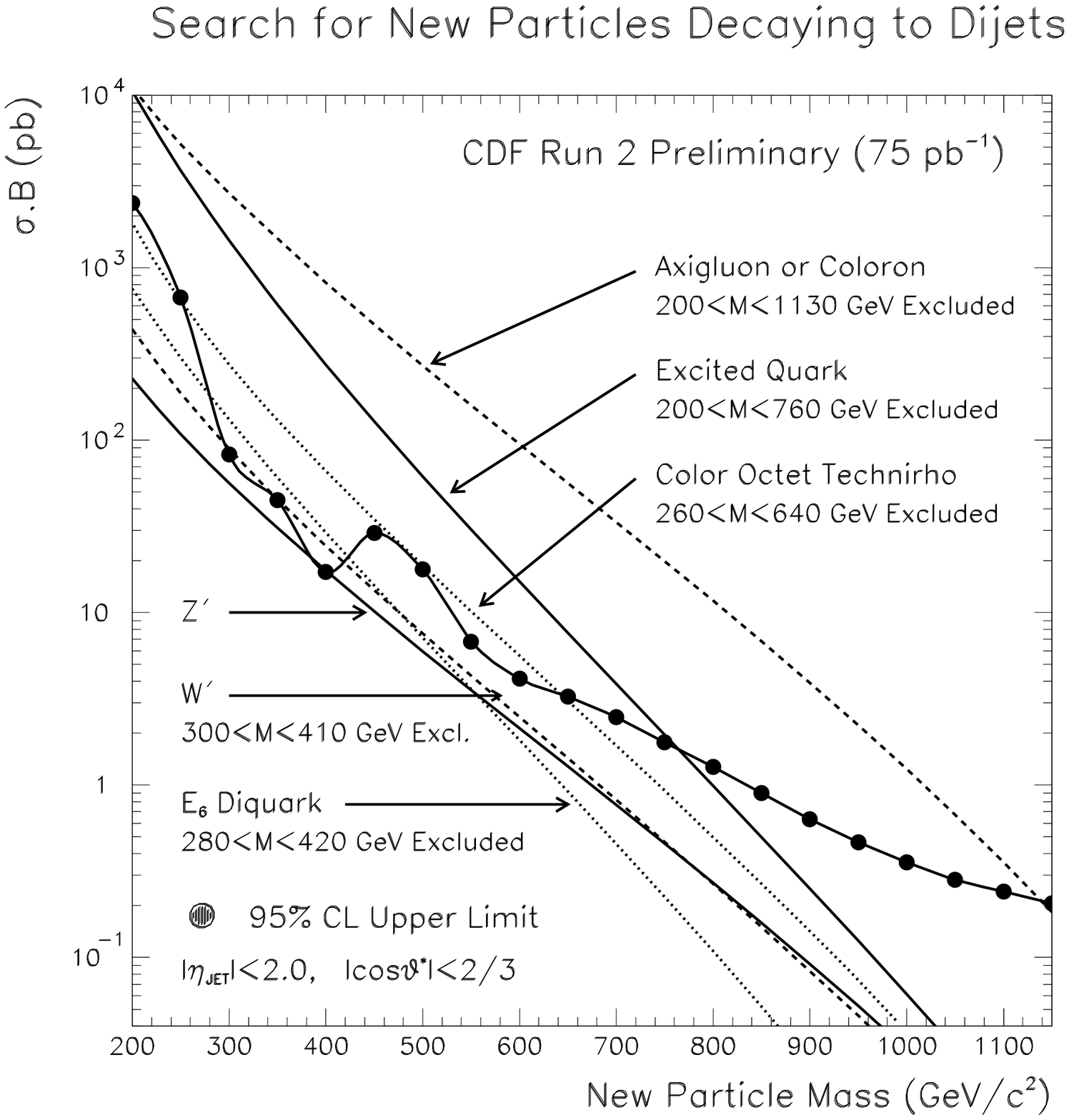}
    }}
  }
  \parbox[t]{.48\textwidth}{\caption{The CDF Run~2 dijet mass spectrum
      divided by the Run~1 spectrum.  \label{cdf_mass1}} }~~~~
  \parbox[t]{.475\textwidth}{\caption{The 95\% CL on the production
      cross section multiplied by $B(X \rightarrow \mbox{dijet})$ and
      acceptance for various theoretical models. 
      \label{cdf_mass2}} 
  }
\end{figure}

 CDF has used a data sample of 75~\ipb\ to carry out a search for new
 resonances95\% confidence level (CL) limits on the production cross
 section in the dijet mass spectrum, and set limits on their
 production within the context of different theoretical models. A
 comparison has been made between the dijet mass spectrum in Run~1 and
 Run~2 (Fig.~\ref{cdf_mass1}) which is in reasonable agreement with a
 LO QCD prediction ({\sc Pythia}). 95\% confidence level (CL) limits
 on the production cross section of new particles are calculated and
 compared to theoretical predictions. The production of excited
 quarks~\cite{excited_quarks} is a convenient model for comparing
 limits between experiments and CDF excludes excited quarks in the
 mass range $200 < M_{W^{\prime}} < 760$ GeV/$c^2$. This compares well
 with the best Run~1 limit of 775~\gevcc\ made by the D\O\
 experiment~\cite{d0_excited}. See Fig.~\ref{cdf_mass2} for limits on
 additional models.

\section{Jet Structure}

 The CDF experiment has made a series of studies of the structure of
 jets and the flow of energy with events with jets in them. These
 measurements allow us to test fragmentation and hadronisation models,
 such as those used in the {\sc Pythia} and {\sc Herwig} Monte Carlo
 simulations. In addition they play an essential role in minimising
 the uncertainties in jet energy scale corrections, and hence are
 essential to measurements of the top mass, for example.

\begin{figure}[hbt]
  \centerline{
    \parbox{.475\textwidth}{\centerline{\includegraphics[width=.475\textwidth ]{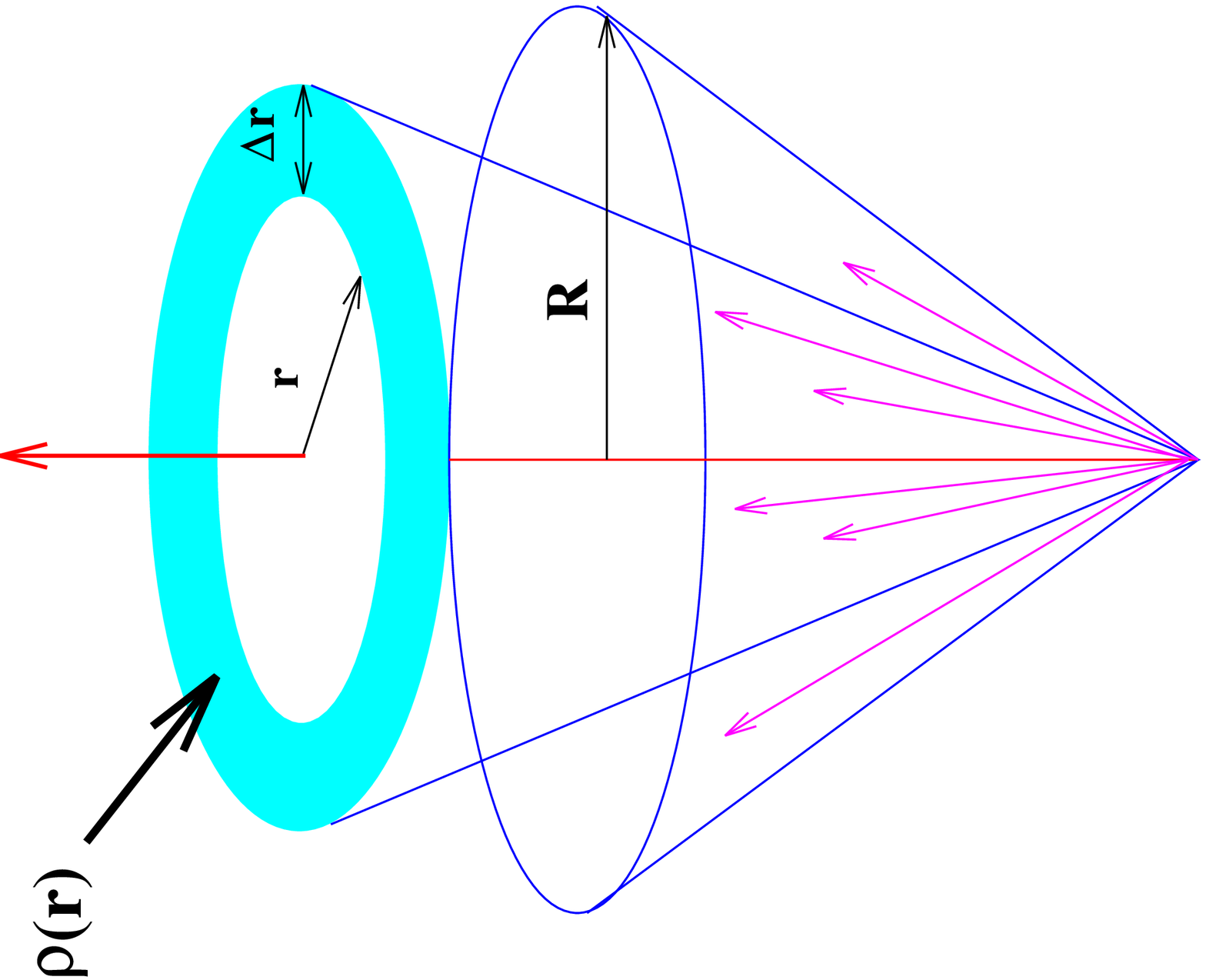}
    }}~~~~
    \parbox{.475\textwidth}{\centerline{\includegraphics[width=.475\textwidth ]{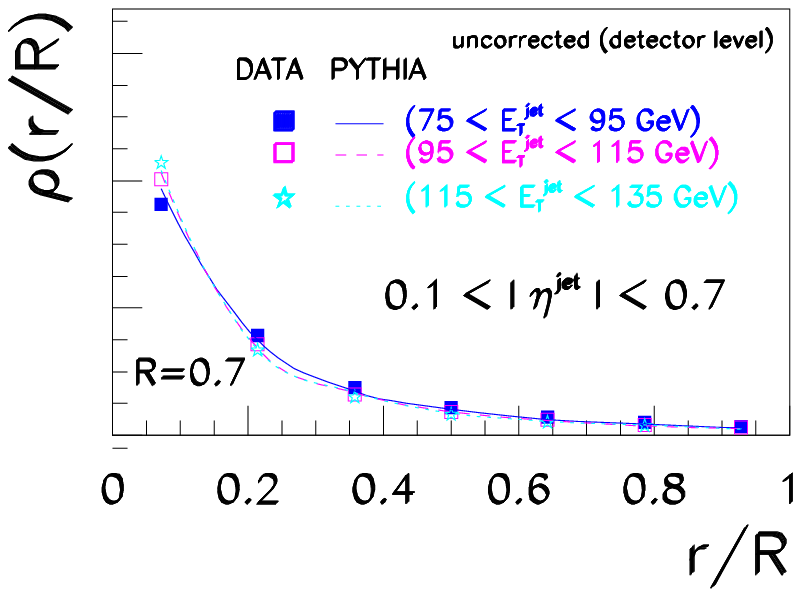}
    }}
  }
  \centerline{
    \parbox[t]{\textwidth}{\caption{The CDF measurement of the jet
    transverse energy shape. The right hand plot illustrates the
    definition of the $\rho$ variable, and the left hand plot shows
    the measurement for $0.1 < \mody 0.7$ with different jet \pt\
    cuts.
    \label{cdf_structure1}} } }
\end{figure}

 CDF has measured the jet transverse energy shape. Jets are identified
 using the cone algorithm with radius ${\cal{R}} = 1.0$. Each jet is
 then divided into 10 sub-cones around the jet axis with sizes varying
 from ${\cal{R}}=0.1$ to ${\cal R}=1.0$, in increments of $\Delta{\cal
 R}=0.1$. The energy in each of these sub-cones is then divided by the
 total energy in the jet (represented by $\rho$,
 Fig.~\ref{cdf_structure1}). Several measurements of $\rho$ are made
 for different \pt\ and rapidity ranges and are compared to {\sc
 Pythia} and {\sc Herwig}. In all cases the simulations are in good
 agreement with the data.

\begin{figure}[hbt]
  \centerline{
    \parbox{.475\textwidth}{\centerline{\includegraphics[width=.475\textwidth ]{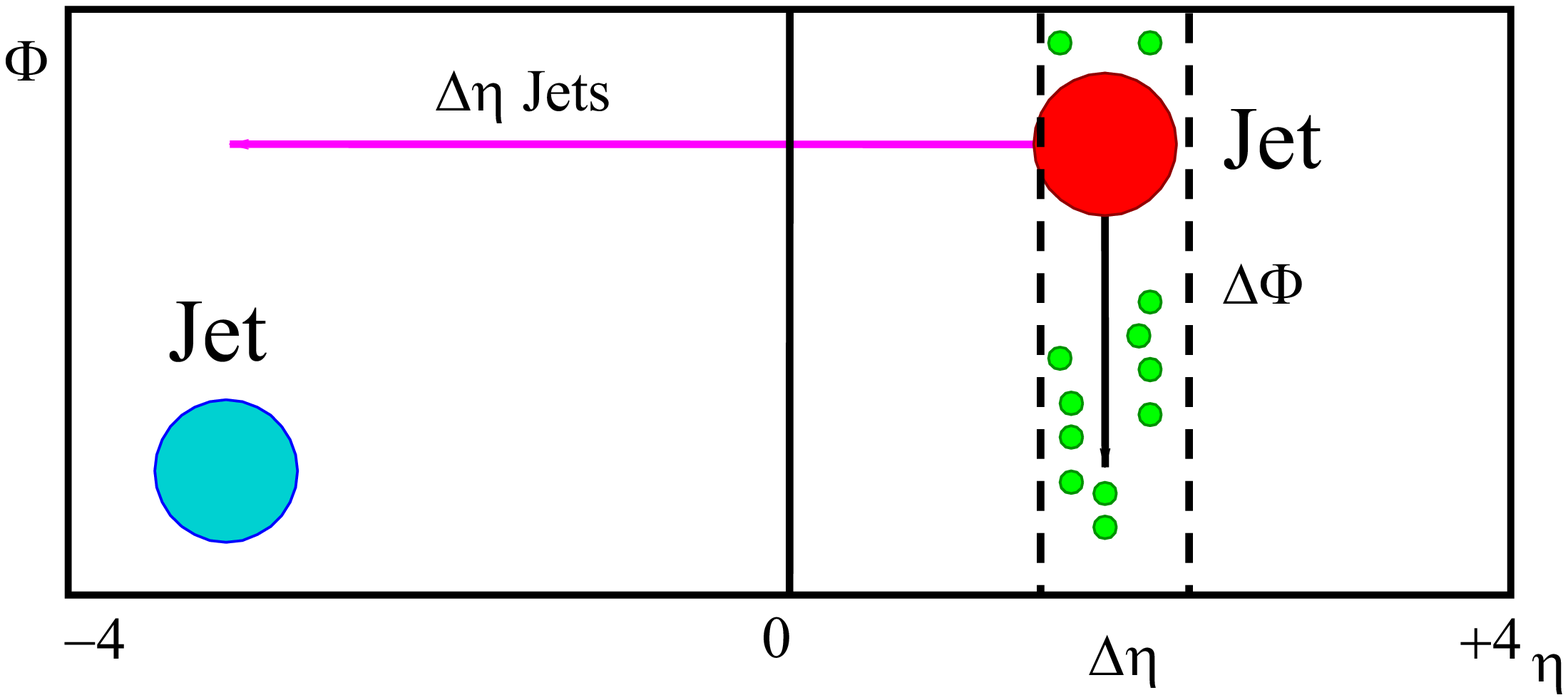}
    }}~~~~
    \parbox{.475\textwidth}{\centerline{\includegraphics[width=.475\textwidth ]{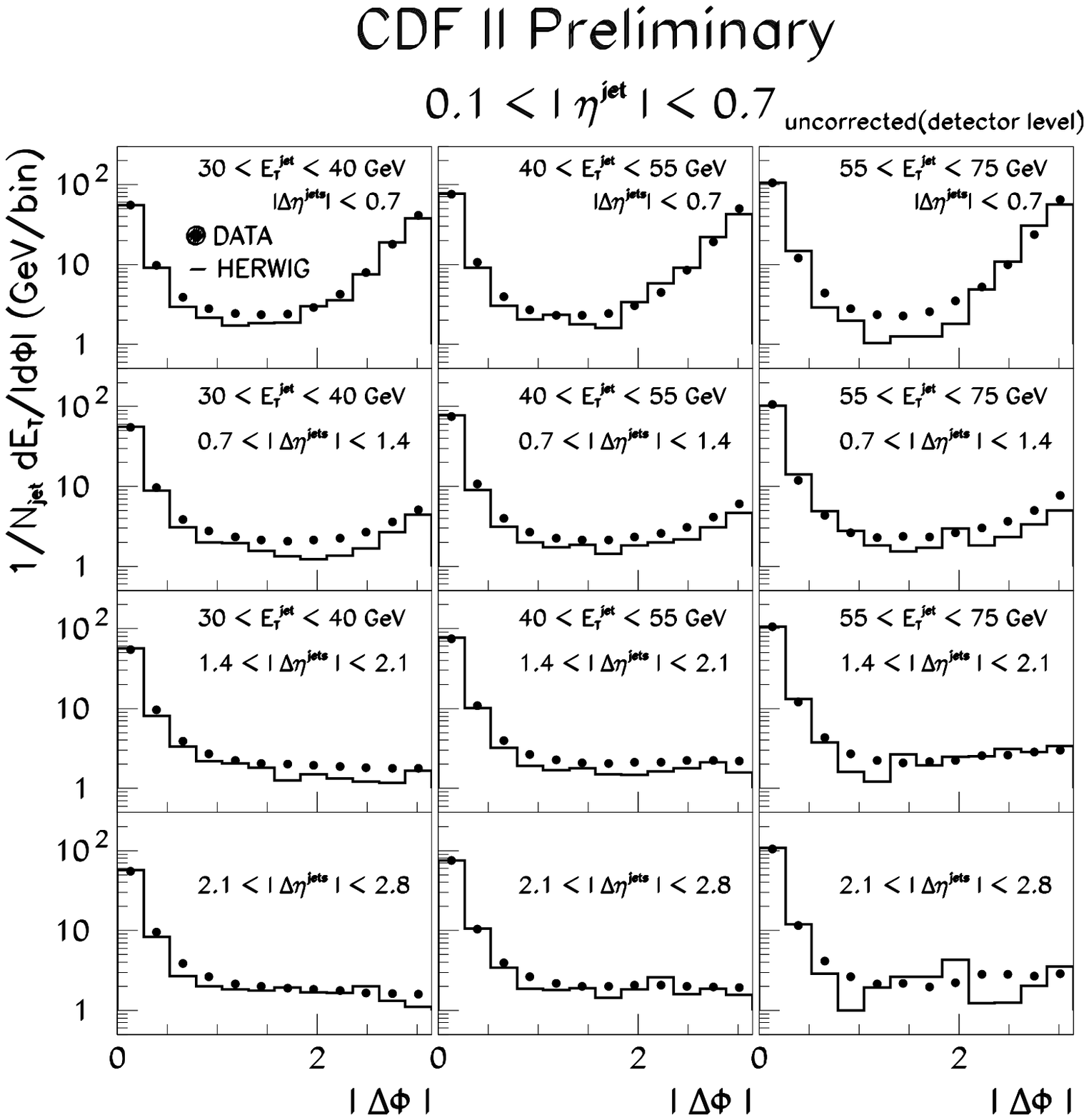}
    }}
  }
  \centerline{
    \parbox[t]{\textwidth}{\caption{ The CDF measures the energy
 outside of a jet, along a band of width $\Delta \phi = 0.7$ centered
 on the $\phi$ axis of the jets as a function of the separation
 between the two leading \pt\ jets in the event, $\Delta \eta$. The
 left hand plot illustrates the measurement and the right hand plot
 compares the data with {\sc Herwig}.
    \label{cdf_structure2}} } }
\end{figure}

 In addition CDF measures the energy outside of a jet, along a band of
 width $\Delta \phi = 0.7$ centered on the $\phi$ axis of the jets as
 a function of the separation between the two leading \pt\ jets in the
 event, $\Delta \eta$. This is an important test of the modeling of
 underlying event in \pbarp\ collisions. The measured data are again
 compared with {\sc Pythia} and {\sc Herwig}
 (Fig.~\ref{cdf_structure2}). Again the data and predictions are in
 reasonable agreement.

\section{Conclusions}

 Both the D\O\ and CDF experiments have made a promising start to Jet
 based analyses using Run~2 data at the Tevatron. I expect that over
 the next year we will see significant improvements in the systematic
 uncertainties, much larger data samples than those collected in
 Run~1. These improvements and the release of new results should lead
 to a much improved understanding of Jet production at hadron
 colliders that will be essential for a good understanding of LHC data
 in the future.

\end{document}